\documentclass[12pt,preprint]{aastex}
\def\plotfiddle#1#2#3#4#5#6#7{\centering \leavevmode
    \vbox to#2{\rule{0pt}{#2}}
    \includegraphics{#1}}

\def\ah{a_{\rm halo}}
\def\nt{N_{\rm test}}
\def\nh{N_{\rm halo}}

\def\al{a_{\rm lump}}
\def\ml{m_{\rm lump}}
\def\nl{N_{\rm lump}}

\def\msun{M_{\odot}}

\begin{document}

\title{
How Lumpy is the Milky Way's Dark Matter Halo?}

\author{
Kathryn V. Johnston\altaffilmark{1},
David N. Spergel\altaffilmark{2} and
Christian Haydn\altaffilmark{1}}

\altaffiltext{1}{Van Vleck Observatory, Wesleyan University,
Middletown, CT 06459 --- kvj@astro.welseyan.edu, ch@astro.wesleyan.edu}
\altaffiltext{2}{Princeton University Observatory, Princeton
University, Princeton, NJ 08544 --- dns@astro.princeton.edu}

\begin{abstract}
        CDM simulations predict that there are hundreds of lumps
of with masses greater than $10^7 M_\odot$ in the Milky Way halo.  
However, we know
of only a dozen dwarf satellites close to this mass.  
Are these lumps simply
lacking in stars or is there a fundamental flaw in our most
popular cosmology?  By studying the tidal debris of known
satellites we can potentially address this question.  In this
paper, we quantify the the effects of the dark matter lumps on 
tidal tails.  The lumps scatter stars in the tidal tails from their
original orbits producing a distinctive signature.  We simulate
debris evolution in smooth and lumpy halos potentials and use our
simulations
to motivate and test a statistical measure of the degree
of scattering apparent in the angular position and radial
velocity measurement of debris stars
--- the ``scattering index''.
We find that the scattering index can in general
distinguish between the levels of substructure
predicted by CDM cosmologies and smooth Milky Way models, but
that the sensitivity of the debris depends on the orientation 
of the parent satellite's orbit relative
to the largest lumps orbits.

We apply our results to the carbon star stream associated with the Sagittarius 
dwarf galaxy (Sgr)
and find that these stars appear to be more scattered than we
expect for debris orbiting in a smooth halo.
However, the degree of scattering is entirely consistent with that
expected due to the influence of the Large Magellanic Cloud, which is on an
orbit that intersects Sgr's own. 
We conclude that the current data
is unable to constrain CDM models. 

Nevertheless, our study suggests that future data sets of debris stars
associated with other Milky Way satellites could provide strong
constraints on CDM models.
\end{abstract}
\keywords{Galaxy: halo --- Galaxy: kinematics and dynamics --- 
Galaxy: structure ---
cosmology: dark matter}

\section{Introduction}

CDM models predict two orders of magnitude more dark matter halos 
than satellites observed around the Milky Way \citep{moore99,klypin99}.
Solutions to this problem include self-interacting dark matter
\citep{spergel00}, truncated
power spectra \citep{kamionkowski00} and 
the restriction of gas accretion to the lowest 
mass dark matter halos to be before the
epoch of reionization \citep{bullock01}.
The first two solutions would get rid of the smallest dark matter halos
entirely, while the latter would predict that only one percent of the 
satellite halos actually contain stars.

In this paper, rather than attempting to solve the problem, 
we discuss how we might tell observationally whether it is 
a problem in the first place.
The satellite dark matter halos in the CDM models comprise about
10\% of the total mass of the parent galaxy on roughly isotropic orbits
distributed throughout the galaxy \citep{font01}.
One idea is to look for signatures of these lumps around
external galaxies in gravitationally lensed images of background quasars
\citep{chiba01,metcalf01}.
We concentrate our study rather closer to home.
If such lumps truly exist around the Milky Way then we would expect them to 
have some dynamical influence on the rest visible Galaxy.
For example, according to \citet{lacey85} if the halo
is entirely made of $10^6 M_\odot$ black holes they would
significantly heat our
Galactic disk.
\citet{font01} used numerical simulations to ask 
whether the coldness of our disk 
could be used in a similar fashion to
limit the distribution of dark matter lumps seen in
cosmological simulations that could be orbiting the Galaxy.
Using one realization of lump masses and orbits taken from a
$\Lambda$CDM model of a Galaxy-sized halo, they found that the heating
caused by the lumps was less than the heating observed in the stellar 
populations in the disk and concluded that the disk is not an efficient
probe of this model.

Streams of debris from the destruction of Galactic satellites are another
example of cold structures within the Milky Way that could be scattered
by substructure in the potential.
These streams tend to align along a single orbit 
\citep{johnston96,johnston98,helmi99} and hence would 
individually have lower cross section to interactions than the Galactic disk.
Nevertheless
we chose to examine the response of these to perturbations rather than the 
disk for several reasons:
they will explore the outer regions of the
Galactic potential that the disk does not experience; we expect other
sources of heating in this region
(structures such as the bar, spiral arms and giant 
molecular clouds in the disk) to be negligible;
if streams from several satellites could be studied, we have the
potential of probing a larger portion of the Galaxy; 
finally, these streams are often even colder
than the disk itself and hence should be more sensitive to scattering.

We approach this problem of scattering numerically rather than analytically
both because of the nature of the predicted CDM mass distribution
of lumps and the low cross sections of the streams.
Scattering by the few most massive lumps in the distribution is expected to
be most important, and the exact degree of scattering will depend on the 
relative orientation of the lump and debris orbits.
Hence the process is dominated by a few strong encounters 
rather than one that can be modeled
(say) by integrating over many weak encounters in the impulsive
regime.
Nor would a simple analytic representation include how the wake in the
halo, excited by the most massive lumps, would affect the stream orbits.
Weinberg's studies \citep{weinberg98a,weinberg95} 
of the affect of the Large Magellanic 
Cloud on the disk of the Milky Way suggest that the inclusion of such
a wake in the halo introduces a large enhancement to the expected
response of the disk.

In \S 2 we present our numerical approach to implementing
the experiments described above. In
\S 3 we compare the evolution of tidal debris in smooth and lumpy potentials
and propose an algorithm for distinguishing between the
two with currently available data.
In \S 4 we apply our results to the stream of
carbon stars known to be associated with
Sgr.
We summarize our conclusions and outline future prospects
in \S 5.

Note that during the final stages of preparation of this manuscript two
other papers have appeared on this subject \citep{mayer01,ibata01}. We discuss
their relation to our own work in \S5.

\section{Methods}

\subsection{Particle Distributions}

\begin{figure}
\plotfiddle{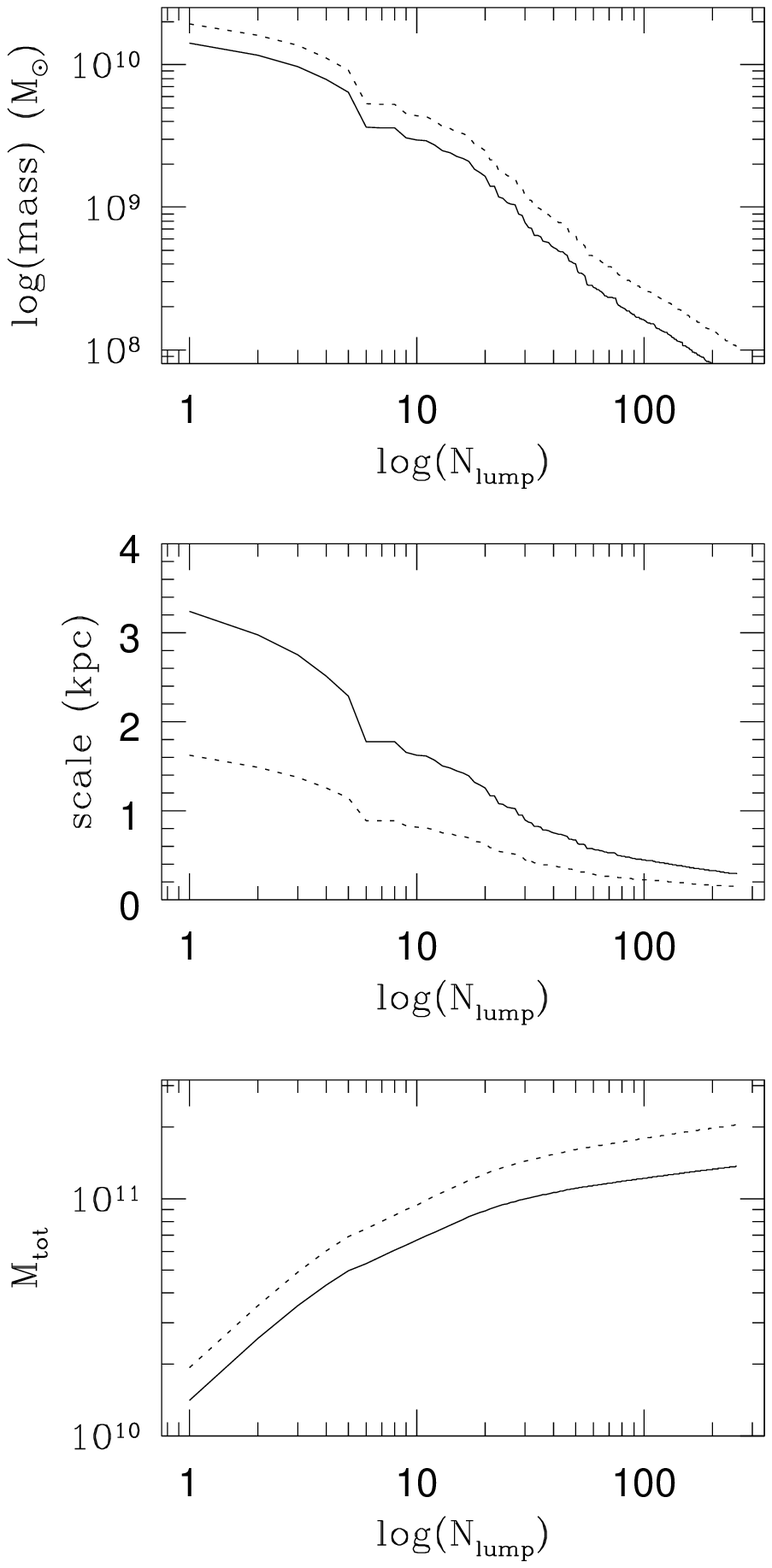}{6.0in}{0}{90}{90}{-170}{-150}
\caption{Masses, scales and cumulative mass of the top 
$\nl$ chosen at random from our $\Lambda$CDM spectrum. 
Solid lines represent parameters for Hernquist models chosen to be
equivalent to the NFW models (dotted lines)
\label{lumps}
}
\end{figure}

We model our system with three sets of particles:
\begin{description}
	\item{$\nh$ {\it smooth halo} particles.}
These are equal mass, and initially distributed as an equilibrium
\citet{hernquist90} model.
This model is assumed to represent a $2.3\times 10^{12}\msun$ 
halo, with scale length
$\ah=30.6$kpc, which mimics the NFW profile \citep{nfw96,nfw97}
of a $v_{200}=200$km/s
halo at $z=0$ in a $\Lambda$CDM Universe (taken from
\cite{navarro00} --- paper since withdrawn, see below).
\citet{navarro00} showed that such halos are incompatible 
with the apparent mass distribution in the Milky Way, but 
we will nevertheless adopt them since we wanted to place our model in a 
consistent $\Lambda$CDM universe.
	\item{$\nl$ {\it lumps}.}
These are each represented by rigid \citet{hernquist90} models, with 
initial positions and velocities chosen at random from the
same \citet{hernquist90} distribution as above.
The masses are
chosen at random from a power-law distribution
$dN/d\ml \propto \ml^{-5.75/3}$, whose form is taken from the end point of
N-body, cosmological simulations of CDM Universes
\citep{klypin99,moore99}.
The upper and lower limits of the distribution are chosen so that
10\% of the parent halo mass is contained in lumps with masses in
the range $7\times 10^7-2\times 10^{10}\msun$ which approximately
matches the level of substructure observed in N-body simulations
\citep{moore99}.
Note that while figures 4 and 5 in
\citet{klypin99} suggest that the degree of substructure
in their $\Lambda$CDM model
is similar to
their CDM case, \citet{font01} found the numbers of satellites in their
$\Lambda$CDM model to be roughly half  those seen in their CDM
model. 
We therefore adopt the \citet{klypin99} and \citet{moore99} level of
substructure as an upper limit to the degree of heating expected in
a $\Lambda$CDM universe.

The scale lengths $\al$ are set to $\al=2 r_s$
and the masses $\ml$ are chosen
so that the potentials mimic 
``Universal'' \citet{nfw96} profiles to within 10\% out to radii $10r_s$,
with the mass-dependence on concentration for
a $\Lambda$CDM Universe taken from \citet{navarro00}.
Note that since completion of the numerical portion of this project the
results of \citet{navarro00} have been withdrawn
and the models we adopted have been shown to be too concentrated.
This will again cause a slight overestimate of the efficiency
of scattering.

\citet{hernquist90} 
profiles were used rather than NFW profiles because of the saving
in cpu-time. 
We have
not tidally truncated the models but 
do not expect this to significantly affect our results.

Figure \ref{lumps} summarizes the mass and scale distribution used
in the simulations with the solid lines representing the parameters for
the Hernquist model and the dotted lines representing the
parameters for the equivalent NFW model.
	\item{$\nt$ {\it test particles}.}
These are massless and are either distributed on circular orbits
at $r=0.5, 1.0, 2.0$ and 4.0$\ah$, or on a range of orbits to
mimic debris streams from Sgr.
The initial conditions for the latter were generated by running a simulations
of satellites of mass $10^8$ 
disrupting in a rigid representation
of the halo potential described above (for simulation technique
see \cite{johnston96}) along an orbit similar to that expected for Sgr (see
\cite{ibata01}).
The time at which particles were lost from the satellite were noted, and 
the positions and velocities of a set of 500 particles used
as initial conditions, starting from a point shortly after the
pericentric passage at which they were first unbound. 
\end{description}

\subsection{Force Computations}

The influence of the smooth halo particles
on each other and all other particles is calculated
using a code that employs basis function expansions to
represent the potential \citep{hernquist92} (we used the MPI version of
the code, adapted from the original by Steinn Sigurdsson and Bob Leary).
This effectively smooths over all strong encounters (which we expect to
be unimportant for this component), but does
follow both large-scale collective fluctuations and the halo's response
to disturbances by lumps.

The influence of the lumps on each other and all other particles
is calculated directly using the analytic form of the \citet{hernquist90} 
potential.

The test particles respond to both the smooth halo and the lumps, but
do not otherwise interact.
This simplification is valid for our application of tidal streams, but
would be a much poorer approximation were we trying to represent a system
such as a disk which has significant self-gravity.

\subsection{Integration}

Simple leap-frog integration was used throughout. 
The major computational expense arose from the large
number ($\nh\ge 10^7$) of smooth halo particles required
(see \S \ref{smooth}).
However, since we were not interested in modeling strong encounters
in this component in detail, we were able to use a large timestep
($dt=0.08$ in simulation units or about 4 Myears) for these particles
corresponding to 1/100th of the dynamical time at the half mass radius
of the parent galaxy.
The lump and test particles were integrated with a much smaller timestep
$dt_{\rm lil}=dt/2^{n_{\rm lil}}$, where $n_{\rm lil}$ was chosen
so the increase in computational cost for these integrations was
not significant.
For $\nh=10^7$, $\nl=20$ and $\nt=1000$, we could take $n_{\rm lil}=8$
(or 256 small steps for every large step and $dt_{\rm lil}\sim 1.6\times 
10^4$ years) with just
a 20\% increase in cputime, and we double checked 
our simulations by rerunning them with $n_{\rm lil}=9$ to confirm
convergence of the results.
In general, we were looking to accurately follow encounters between
features of a kpc with
lumps traveling at 200km/s  
so we would expect $dt_{\rm lil}=(1{\rm kpc}/200{\rm km/s})/100 \sim
50000$ years to be a 
sufficiently small integration time-step.

\subsection{Initialization}

The smooth halo distributions were allowed to run for ten dynamical
times to erase any effects from generating the initial conditions.
For each set of lumps, the combined smooth halo and lumps
were run for an additional ten dynamical times as the lumps were
grown slowly from zero to full strength to ensure no effects from
suddenly introducing them.
Finally, the output of the latter simulations were used as initial
conditions in which the test particles were also run.

\subsection{Required Particle Number, Number of Simulations and CPU-cost}
\label{smooth}

\begin{figure}
\plotfiddle{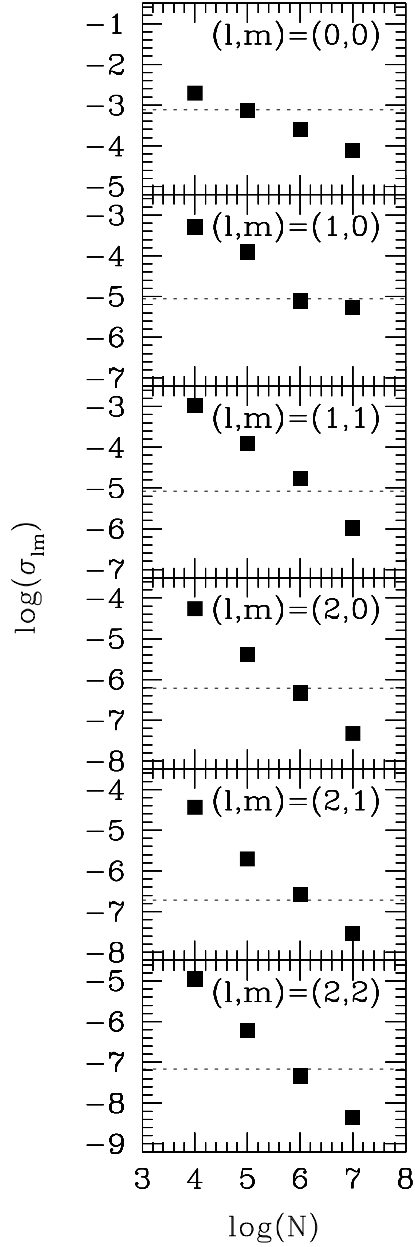}{6.0in}{0}{90}{90}{-320}{-150}
\caption{Dispersion (calculated over 10 dynamical times)
in the potential energy associated with
each $(l,m)$ spherical harmonic in a halo realized by 
$\nh$ particles.
The dashed line shows the same calculation for a halo with $N=10^7$
particles with the most massive CDM lump orbiting in it.
\label{spotlm}
}
\end{figure}

The finite number of halo particles will introduce potential fluctuations
in addition to those due to the lumps \citep{weinberg93,weinberg98b}
so we need to ensure that the dominant cause of test-particle 
scattering is due to the lumps and not to numerical noise.
To quantify the level of fluctuations in the halo due to the
finite number of particles we first ran models with $\log\nh =4,5,6$ and 7
in isolation for ten dynamical times and recorded the 
basis-function
expansion coefficients $A_{nlm}$ calculated by the
code during this time (where $n$ refers to
the radial basis function and $(lm)$ refers to the spherical harmonic ---
see \citet{hernquist92}). 
The total potential energy can be written in terms of this expansion 
as $\Phi =\Sigma_{nlm} A_{nlm}^2$.
Hence the potential energy associated with each of
the spherical harmonics is $\Phi_{lm}=\Sigma_n A_{nlm}^2$.
The points in
Figure \ref{spotlm} show the dispersion in the 
lowest order $\Phi_{lm}$ recorded 
during the isolated simulations.
The dotted line shows the same measurement made in a simulation
containing $\nh=10^7$ particles  and just one NFW lump.
The plot suggests that to correctly model the influence of lumps 
on tidal streamers requires $\nh \sim 10^7$ particles since otherwise
scattering due to the finite resolution of the halo will compete
with the effect being measured.

A second numerical consideration is that, so long as the most massive 
particles have the greatest influence, the size of the effect we are
trying to measure will depend on the exact choice of lump and streamer
orbits. To gain a fair impression of the general evolution it is
necessary to run many simulations with different lump orbits.

The combined requirements of large particle number and multiple 
runs supports our choice of running idealized simulations of 
fully formed galaxies over a restricted amount of time rather than
trying to perform this experiment fully self-consistently in
a cosmological context, which would add additional computational 
expense.
Each simulation was first ``relaxed'' for 500 large steps and
then followed for 1000 large time steps, or
ten dynamical times (corresponding to about 4 Gyears)
for a total cpu-cost of $\sim 225 (\nh/10^7)$hours.
We ran more than 50 such simulations (a subset of which are presented
here).
To reduce this to manageable proportions (in memory and time),
simulations were run on multiple nodes of the Wesleyan Beowulf-class 
supercomputer (WesWulf).

\subsection{Parameter Sets}

We present the results from a total of 37 simulations in this paper.
Each simulation contained $\nh=10^7$ halo particles and
$\nt=4000$ test particles.
The ``control'' simulation contained no halo lumps.
The remaining 36 consisted of 6 sets of 6 random realizations of
lump orbits for $\nl=1,4,16,64,128$ and 256. Each set of simulations
contained the same lump mass distribution, with increasing numbers of
lumps exploring further down the mass function.
Since test particles were distributed on orthogonal orbits
our final results contained 12 different realizations of debris
distributions, all starting from the same initial conditions but
experiencing different time dependent potentials.

\section{Results}

\subsection{Evolution of Orbits in Lumpy Potentials}

\begin{figure}
\plotfiddle{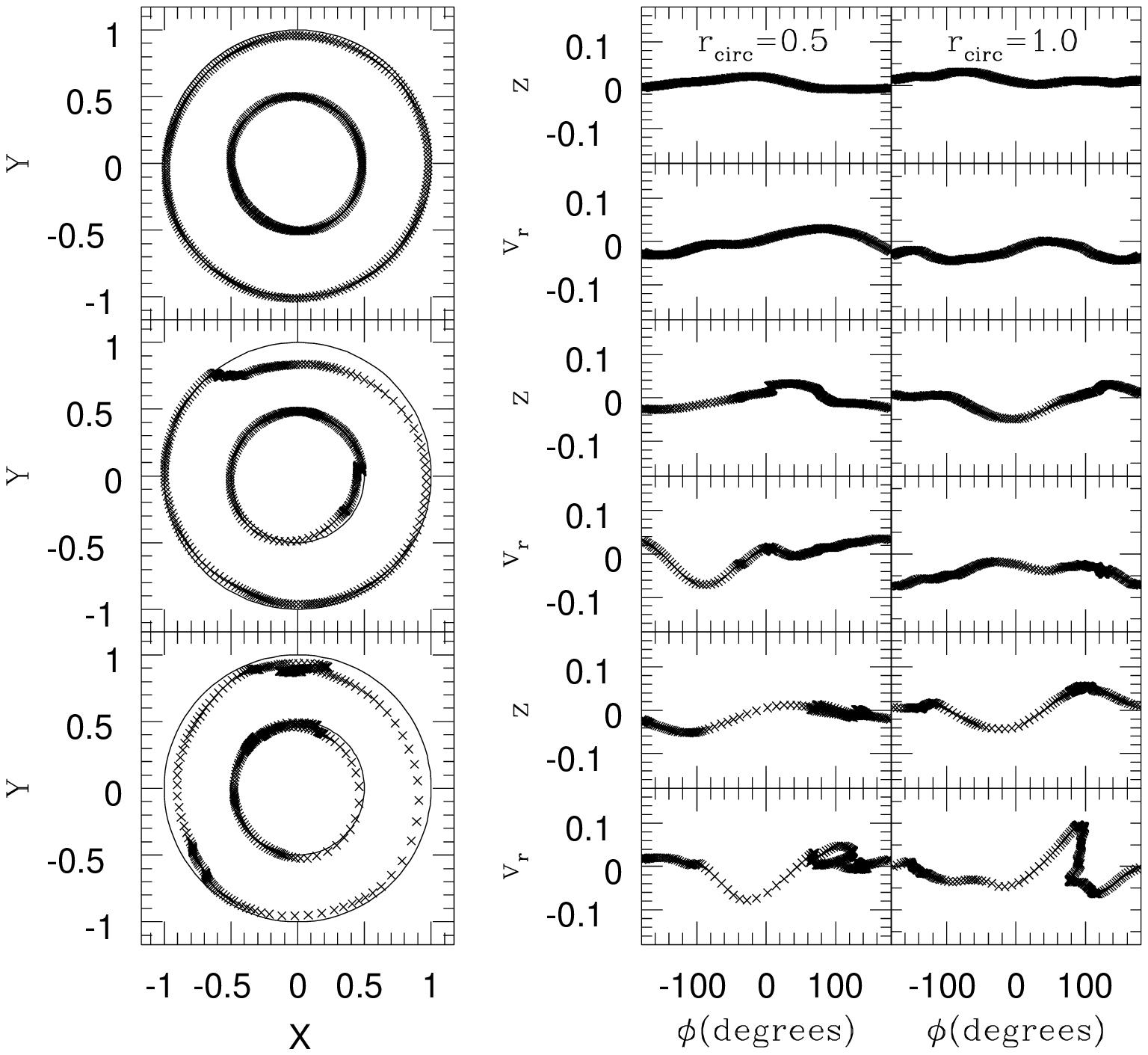}{6.0in}{0}{80}{80}{-200}{-100}
\caption{
Final positions (left hand panels) and ``observations'' (right
hand panels) of test particles initially on circular orbits after 
1.3 (top panels), 2.6 (middle panels)
and 4 (bottom panels)Gyears of evolution in a halo containing $N=10^7$ particles and
$\nl$=256 lumps.
\label{pscirc}
}
\end{figure}

As an example of evolution in a lumpy potential
Figure \ref{pscirc} plots the phase-space distribution of 
particles initially on circular orbits at $r=0.5$ and 1.0 kpc
at times $t=0.8, 2.4 $ and 4Gyrs
from one simulation containing $\nl=256$ lumps.
The left hand panels show positions in the orbital plane
while the right hand panels show what observations we might
make of the stars
(assuming we could identify the initial orbital plane of the debris):
angular distance $d\theta$
from the orbital plane and line-of-sight velocities $v$
as a function of angle $\Psi$ along the debris.
The two things immediately apparent are: (i) 
the orbital plane precesses 
over time (visible as a sinusoidal shape in the
lowest $d\theta$ {\it vs} $\Psi$ panel); 
and (ii) particles initially distributed smoothly along the orbit become
bunched in angular position and velocity.

\begin{figure}
\plotfiddle{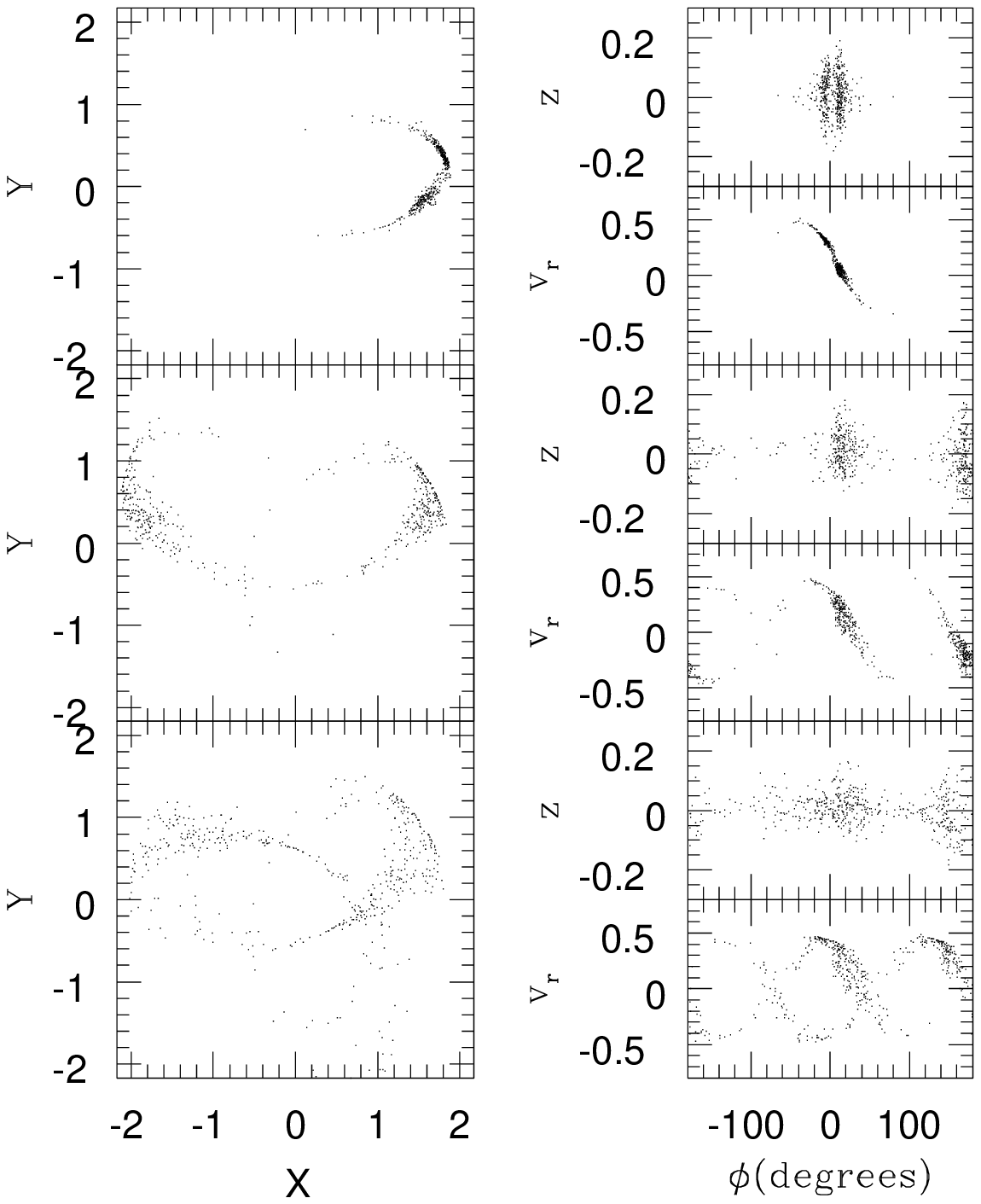}{6.0in}{0}{80}{80}{-200}{-100}
\caption{
Final positions (left hand panels) and ``observations'' (right
hand panels) of test particles initially on Sgr debris orbits after
1.3 (top panels), 2.6 (middle panels)
and 4  (bottom panels) 
Gyears of evolution in a halo containing $N=10^7$ particles and
$\nl$=0 lumps.
\label{pssgr}
}
\end{figure}

\begin{figure}
\plotfiddle{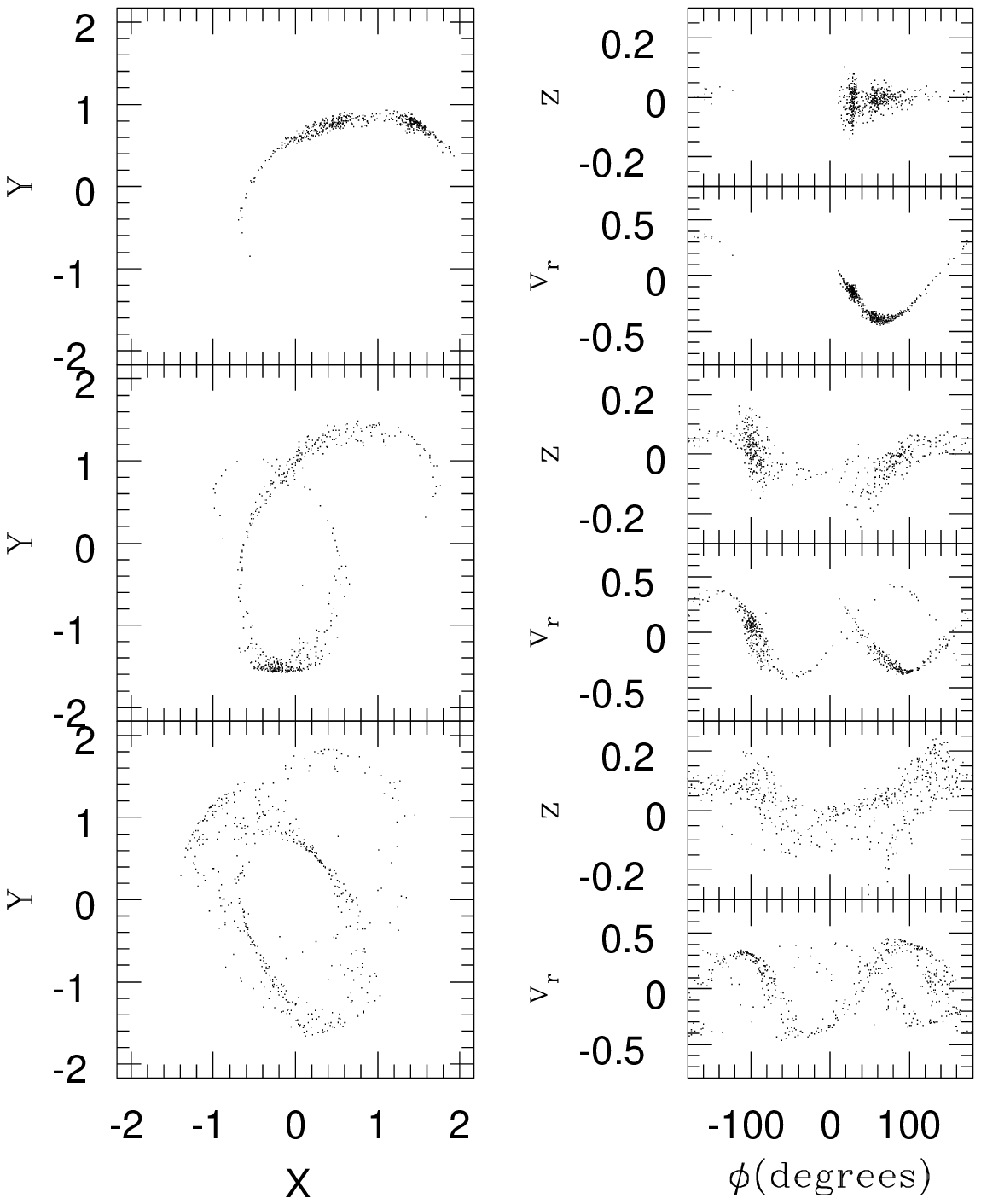}{6.0in}{0}{80}{80}{-200}{-100}
\caption{
Final positions (left hand panels) and ``observations'' (right
hand panels) of test particles initially on Sgr debris orbits after
1.3 (top panels), 2.6 (middle panels)
and 4 (bottom panels)
Gyears of evolution in a halo containing $N=10^7$ particles and
$\nl$=256 lumps.
\label{pssgrl256}
}
\end{figure}

Figures \ref{pssgr} and \ref{pssgrl256} repeat the above plots for the
tidal debris particles in simulations with zero and $\nl=256$ lumps.
Comparison of the right hand panels of these plots clearly show the
orbital precession described above in the lumpy case. 
There is also some indication of additional structure along the orbit, but
this is somewhat hidden by the intrinsic non-uniform distribution of
debris along the orbit.

\subsection{Interpreting Observations}

From our qualitative assessment of Figures \ref{pscirc}-\ref{pssgrl256}
we know we need to design a statistic that is sensitive to scattering
in both angle and velocity 
along a debris stream rather than to large scale effects such as
overall heating. (Although precession is an additional affect it is not
observable since do not in general know the original orbital plane
of the satellite.)
A natural choice is to look at fourier series in the observed quantities:
\begin{eqnarray}
	B_{\Psi,m}&=&|\Sigma_k d\theta_k \exp^{i m \Psi_k}| \cr
	B_{v, m} &=&|\Sigma_k d\theta_k \exp^{i m v_k/v_{\rm max}}|. 
\end{eqnarray}
In general we expect the low order $B_m$ to contain signal that could be due 
to periodic nature of debris orbits (in $v$) or debris density (in $\Psi$)
along the orbit. However, higher $B_m$ should be sensitive to scattering
in the debris distribution.

To mimic how this might be applied to a real data set (such as Sgr) 
we: \\	
(i) ``Observe'' the Galactic latitude and longitude and line-of-sight 
velocity $v$ of  $n$ debris particles in our simulations from a
viewpoint 8kpc from the Galactic center; \\
(ii) Define the orbital plane to be the best fit great circle to the
angular data \citep{johnston96}; \\
(iii) Find $\Psi$ and $d\theta$ relative to this great circle; \\
(iv) Define a ``scattering index'' by summing over the fourier
expansion -
\begin{equation}
\label{bbb}
	B=\sqrt{\Sigma_{m=5,10} B^2_m}
\end{equation}
where
\begin{equation}
\label{bm}
	B^2_m=B_{\Psi,m}^2+B_{v,m}^2.
\end{equation}
Note that the limits in the summation were specifically chosen so that the
statistic is not sensitive to large scale effects such as 
intrinsic non-uniformity
of tidal debris.

The top panel of Figure \ref{b500} plots $B_m$ against $m$ calculated from 
all 500 of the debris particles from the $10^8 M_\odot$ satellite
at the end of the simulations which each
contained 256 lumps (open squares and solid lines)
and contrasts these with
the simulation containing no lumps (closed squares and bold lines).
The bottom panel plots the scattering index for all the simulations
as a function of number of lumps (open squares) with the bold line showing
$B$ for the no-lumps case.
In general, we find we expect to be able to use this index to distinguish
between s$\Lambda$CDM and non-lumpy Milky Way halos roughly ninety 
percent of the time.
There is a tendency for the majority of the scattering to be caused by
the very largest lumps.
However, the large variance in the results for a given $\nl$ emphasizes 
that the degree of scattering is very sensitive to the relative 
orientation of lump and debris orbits.

\section{Application to Observations: Sagittarius' Carbon Star Stream}

\citet{ibata01} report the discovery of a set of carbon stars which closely
align with the great circle defined by Sgr's position and proper motion.
They find that this stream is too thin to be easily explained in models of the 
Galaxy that have oblate halos because debris
orbits in non-spherical potentials precess beyond the 
width of the stream within the typical lifetimes of the Carbon stars
(47 of the 104 stars lie within 10 degrees of Sgr's great circle which 
intercepts less than one quarter of the total survey area).
Such a cold stream provides an obvious testing ground for models that
predict dark matter substructure around the Milky Way.

Carbon stars have ages ranging from a few to 6 or 7 Gyears.
Hence we expect the length of our simulations (4 Gyears) to be
a fair representation of how long this debris may have been
orbiting independently.

\begin{figure}
\plotfiddle{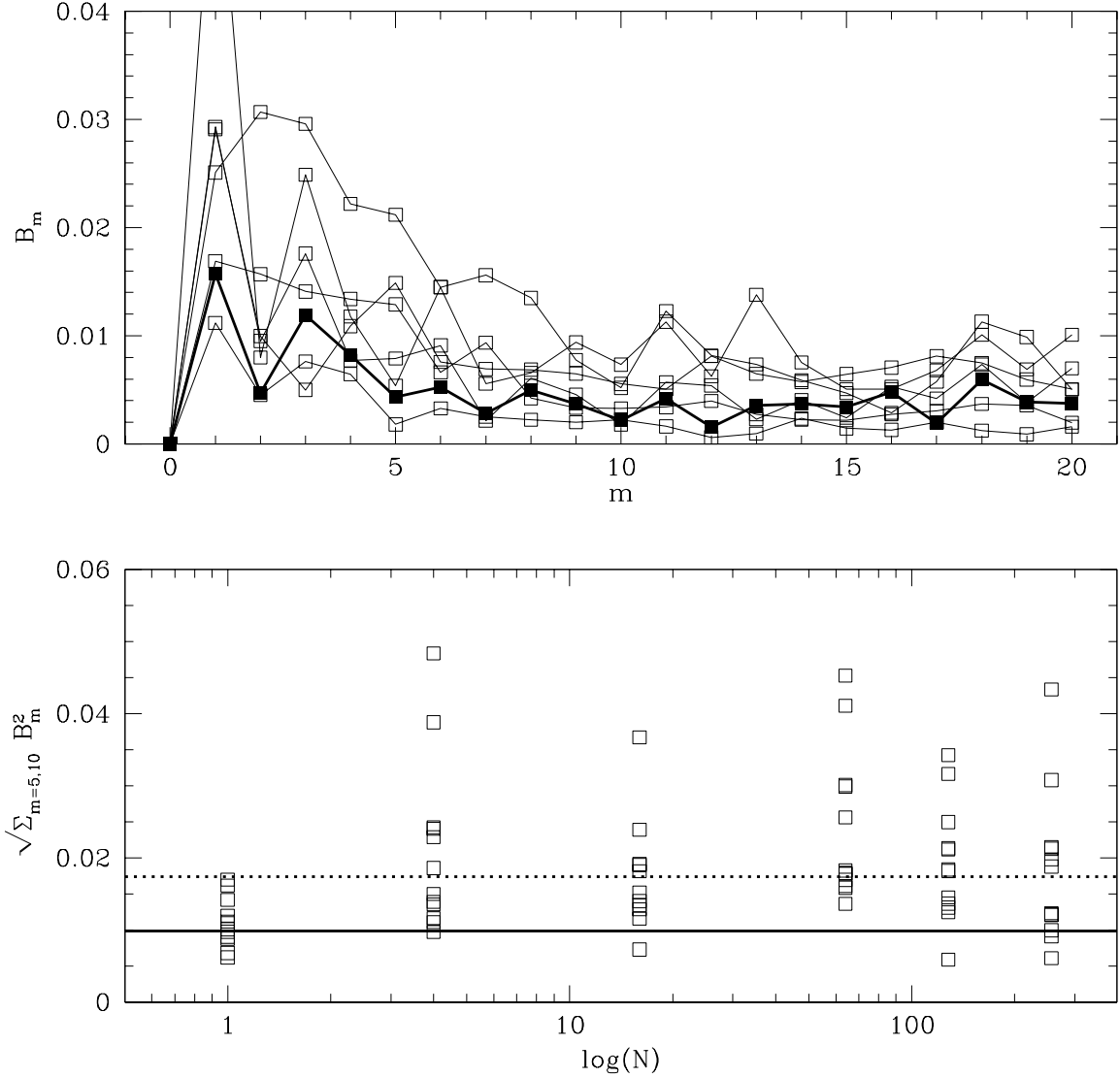}{6.0in}{0}{80}{80}{-240}{-100}
\caption{
Upper panel shows
components of the scattering index $B_m$, calculated from the final
observations of 500 Sgr debris particles, plotted as a function of fourier
number $m$ (see equation [\ref{bm}] for definition). Open squares and
solid lines show results for all realizations of halos with $\nl=256$
lumps on random orbits. Filled squares and bold lines show the result
for evolution in a smooth halo.
Lower panel shows the scattering index $B$ (see equation [\ref{bbb}]) 
calculated
for all 500 debris particles at the end of each simulation, as a function of
the number lumps in the simulation. The bold line indicates $B$ for a
simulation containing no lumps and the dotted line is for one containing 
a single lump of mass $10^{10}M_\odot$ on an LMC-like orbit.
\label{b500}
}
\end{figure}

\begin{figure}
\plotfiddle{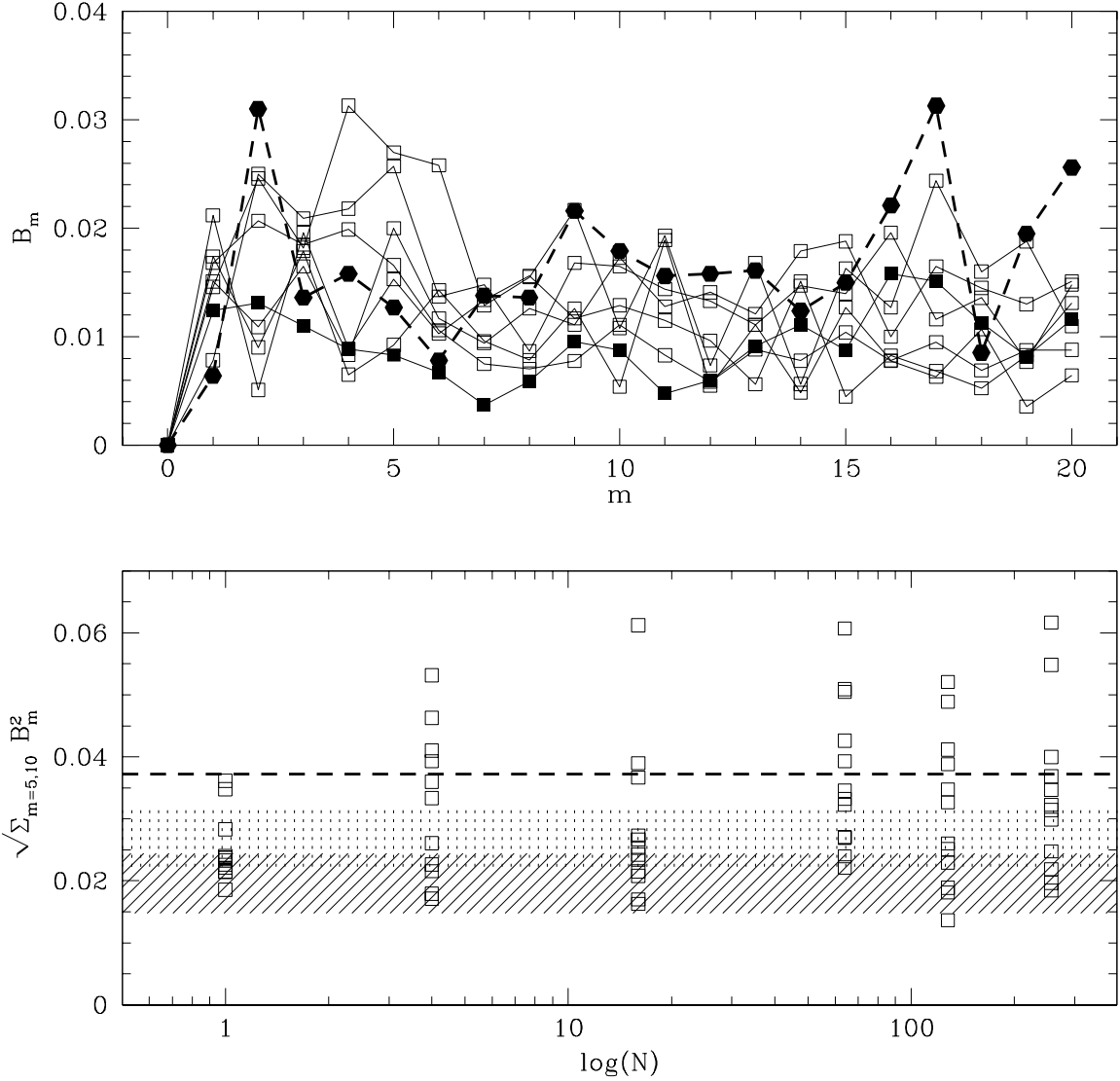}{6.0in}{0}{80}{80}{-240}{-100}
\caption{As Figure \ref{b500} but for 47 particles selected to be within 
10 degrees of the best fit Great Circle to the debris particles at the end
of the simulation, and at Galactic latitudes $|b|>30$ degrees.
The bold dashed lines in both panels 
shows the result of applying the same statistic to the
47 Carbon stars found within 10 degrees of Sgr's orbital plane. 
In the lower panel, the lower
shaded region indicates the dispersion around the mean for 10 
sets of 47 randomly selected particles from the simulation containing no
lumps.
The upper shaded region shows the same region for a simulation containing just
one lump, mass $10^{10}M_\odot$ on an LMC-like orbit.
\label{b47}
}
\end{figure}

Figure \ref{b47} repeats Figure \ref{b500} but for 47 tidal
debris particles chosen at random
at the end of the simulations.
In this case the particles were again viewed at point 8kpc from the
Galactic center chosen slightly out of the orbital plane of the satellite
to reflect the Sun's orientation relative to Sgr's own orbit.
To mimic the real survey, we also
restricted our simulated survey to only those particles further than
30 degrees from the Galactic disk and within 10 degrees of the
best-fit great circle of the full debris distribution.
In the upper panels of Figure \ref{b47}
the points and lines represent the results from just
one set of particles in each simulation that contained $\nl$=256.
In the lower panels
the points represent the results for one set of 47 particles chosen
from each simulation.
The lower shaded region in the lower panels shows dispersion around
the average of the results for ten different
particles sets in the no-lumps case.
The bold dashed line in both panels is the result of the identical analysis
performed on the Sgr carbon star data set.
The figure suggests that the Sgr carbon star set exhibits 
a level of substructure inconsistent with debris orbits on a smooth, spherical
potential.
 
The ``lumps'' that we are already aware of in the Milky Way's halo consist
of the Large and Small Magellanic Clouds 
and the eight other dwarf spheroidal 
satellites.
Since our results clearly depend on the exact orientation of the orbits of the 
largest lumps masses relative to the debris orbits we ran one final
simulations in which we integrated the Sgr debris in a halo containing
a satellite of mass  $10^{10}M_\odot$ 
in an orbit like that of the LMC
(roughly perpendicular to Sgr's orbit with a pericenter comparable to 
Sgr's apocenter). 
The results of the application of the scattering index to 
to the output from the simulation are shown as bold dotted lines in
Figures \ref{b500} and \ref{b47}. We conclude that for the current Sgr
data set
the degree of scattering is entirely consistent with debris perturbed
by the LMC alone. Moreover, Figure \ref{b500} implies that
even with a larger data set associated with
Sgr it will be
difficult to distinguish between a s$\Lambda$CDM halo and a smooth
one.

\section{Conclusions and Future Prospects}

In this paper we showed that it is possible to distinguish between 
smooth and lumpy Milky Way halos by quantifying the coldness of tidal
streams. We proposed a ``scattering index'', based on position and
radial velocity measurements of stars, that is sensitive to small
scale perturbations in the debris rather than large scale effects 
such as variations in debris density or line of sight
velocity (due to its basic dynamical properties).
We found that this statistic, when applied to
measurements of 500 stars in a single debris trail,
could distinguish between smooth, spherical
Milky Way
models and those containing a level of substructure consistent with
s$\Lambda$CDM models 90\% of the time.
Most scattering was due to the few largest lumps and the degree of
scattering was very sensitive to the exact orientation of lump and debris
orbits.

These results agree qualitatively with those of \citet{mayer01} and 
\citet{ibata01b}.
\citet{mayer01} present images from simulations of tidal tails evolved
in a fully self-consistent s$\Lambda$CDM models to illustrate to what extent
the tails are disrupted. This approach has the
advantage of being able to follow the evolution of debris within a
cosmological context, but (as noted in \S2.5), is limited by the resolution
and cost of such simulations to single realizations of galaxies that
may not be adequately resolved to conquer intrinsic scattering due to
numerical noise.
\citet{ibata01} simplify our own approach by modeling only the
influence of lumps on the debris, not accounting for the
halo wake in their simulations and hence (as they note) underestimating the
true magnitude of the lump influence.
Nevertheless, they demonstrate that scattering due
to dark matter lumps should be easily detectable in the angular
momentum distribution of debris from globular clusters --- an experiment
that will become feasible with the launch of the GAIA satellite in the second
decade of the twenty first century.

In contrast to \citet{ibata01b}, our own aim was to look at the
feasibility of constraining the halo dark matter distribution with data
available today or in the near future. 
Following this aim 
we applied our scattering index to the carbon star stream associated with
Sgr and found that it contains more scattering than would be expected
if it were orbiting in a smooth Milky Way. 
We then demonstrated that the level of scattering in Sgr's debris is
entirely consistent with perturbations by the LMC alone.
We conclude from our study
that the current data is unable to place limits on the
level of dark matter substructure in the Milky Way's halo.
Moreover, the specific alignments of the LMC's and Sgr's orbits
means that a larger sample of Sgr's debris is unlikely to improve the
sensitivity of our statistic.

Two other factors could be responsible for 
the large $B$ measured for the Sgr sample:
we have not taken into account the contamination of the carbon star
sample by non-Sgr stars; and we have assumed that the Galaxy is 
perfectly spherical and that the stream is not broadened by precession
of the debris orbits.
Despite these limitations we remain optimistic about the use of tidal debris
to constrain dark matter substructure in the future.
The former problem should be solved once we have a large enough data set
to identify Sgr debris by continuous velocity variations across the sky, and
the latter could be addressed with a clean sample by looking for anomalous
local scatterings in velocity.

In addition, 
there is mounting evidence to suggest that many others of the Milky Way's
satellites (and globular clusters) have associated debris 
\citep{grillmair95,irwin95,majewski00}.
This evidence is currently limited to overdensities of material close to these
objects and hence too young dynamically to be of use for our study.
However several deep-halo surveys are currently underway that have the
potential of tracing these streams out further \citep{sdss1,sdss2,ggss,spag}.
If extensive debris is found then these satellites,
all of which have lower masses than Sgr, would provide even colder
streams which should be even more 
sensitive probes of substructure. 
They are also on different orbits, exploring other regions of the halo, where
the LMC's influence will be less dominant.
Looking even further to the future, astrometric satellite programs such as
NASA's SIM and ESA's GAIA will provide the two additional dimensions of 
proper motion to allow a more rigorous measure of scattering
to be defined, such as that proposed by \citet{ibata01b}.

\acknowledgements
We would like to thank Mike Irwin for generously making the Sgr carbon star
data set available to us, Steinn Sigurdsson for sending us the
MPI version of the SCF code and Ben Moore and Julio Navarro for helpful
comments on the s$\Lambda$CDM model used.
KVJ was supported by  NASA LTSA grant NAG5-9064 and
CH was supported by funds from Wesleyan University.
DNS was partially supported by NASA ATP grant NAG5-7154.

\clearpage

\end{document}